\documentstyle[12pt]{article}

\newcommand{\be}{\begin{eqnarray}}
\newcommand{\ee}{\end{eqnarray}}

\newcommand{\aovb}{\frac{\pi^2}{6}|\langle\bar qq\rangle|\tau^3}
\newcommand{\asovbs}{\frac{\pi^4}{36}|\langle\bar qq\rangle|^2\tau^6}
\newcommand{\aqovbq}{\frac{\pi^6}{216}|\langle\bar qq\rangle|^3\tau^9}
\begin{document}
\pagestyle{empty}
%vbox to  0.8in{}

\vfill
\eject
\begin{flushright}
SUNY-NTG-92/45
\end{flushright}

\vskip 1.5cm
\centerline{\bf\LARGE Baryonic Correlators}\vskip0.5cm
\centerline{\bf\LARGE in the Random Instanton Vacuum}
\vskip 1.5cm
\centerline{ T.~Sch\"afer\footnote{supported in part by the
Alexander von Humboldt foundation.}, E.V.~Shuryak and J.J.M.~Verbaarschot}
\vskip .6cm
\centerline{\it Department of Physics}
\centerline{\it State University of New York at Stony Brook}
\centerline{\it Stony Brook, New York 11794, USA}
\vskip 2cm

\centerline{\bf Abstract}
\vskip 0.3cm
  This is the third paper of a series devoted to a systematic study of
QCD correlation functions in the framework of an instanton model for the
QCD vacuum. In this paper we concentrate on the quark-quark (diquark) and
three quark (baryon) channels. We have found that the quark-quark interaction
resembles the one between quarks  and antiquarks. Similar to the pion and
rho channels, the interaction in the scalar isospin $I=0$  and vector $I=1$
diquark channels is completely different: the former has a much stronger
attractive interaction.  As a consequence, the $SU(3)$  octet (nucleon) and
decuplet (delta) correlators are also found to be qualitatively different.
Using a complete set of all available correlation functions, we determine
masses and coupling constants for the nucleon and delta. Our results
agree surprisingly well with the first lattice data on point to
point correlators.

\vfill
\noindent
\begin{flushleft}
SUNY-NTG-92/45\\
April 1993
\end{flushleft}
\eject
\pagestyle{plain}

\eject
\pagestyle{plain}
\renewcommand{\thefootnote}{\arabic{footnote}}
\setcounter{footnote}{0}
\setcounter{page}{1}

\setlength{\baselineskip}{23pt}
\vskip 1.5 cm
\renewcommand{\theequation}{1.\arabic{equation}}
\setcounter{equation}{0}
\centerline{\bf 1. Introduction  }
\vskip 0.5 cm
  This is the third paper of a series devoted to a detailed investigation
of the QCD point-to-point correlation functions\footnote{A detailed
review on QCD correlation functions recently appeared in \cite{Shuryak_cor}.}
in the framework of the Random Instanton Liquid Model (below RILM). After
considering the propagation of quarks \cite{Shuryak_Ver_il1} and
quark-antiquark (mesonic) states \cite{Shuryak_Ver_il2} we now
consider the baryon sector.

  This model of the QCD vacuum was proposed by one of us
\cite{Shuryak_1982} in 1982. Using a variety of arguments, it was
suggested  that the density of 'tunneling events'
(instantons and antiinstantons combined) is about $n_0=1\, fm^{-4}$,
while their typical size is $\rho_0=1/3\, fm$. These two
numbers were shown to reproduce two global properties of the QCD vacuum,
the gluon and quark condensates.

The investigation of correlation functions in the RILM
was started in reference \cite{Shuryak_A4} for the pseudoscalar channels.
Not only does the model explain why the OPE-based sum rules fail in
the case of pseudoscalars \cite{Geshkenbein_Ioffe,Novikov_etal}, but
it was also shown to reproduce deviations from  asymptotic
freedom at small distances in a quantitative way.

  During the past decade significant efforts were devoted to
derive the instanton picture of the QCD vacuum from first
principles. First variational \cite{Diakonov_Petrov} and later numerical
\cite{Shuryak_1988,NVZ} methods were applied to this problem, with the hope
to develop a  self-consistent theory of interacting instantons. In spite of
these efforts, a clear
cut derivation  from first principles is still missing.
  Recently, we have performed further studies along these lines,
substituting  the rather arbitrary trial functions used previously
by the 'stream-line' set of configurations found in \cite{Verbaarschot}.
These studies have shown that we still lack an understanding of how exactly
one should treat 'a bottom of the valley', the  configurations with a
very close instanton-anti-instanton pair.
We are now preparing a separate publication, in which the current
understanding of this problem will be discussed.

In the RILM we simply assume that except for the size, which we keep fixed,
the distribution of the collective coordinates associated with the instantons
and antiinstantons is completely random. We consider the  RILM as  the simplest
possible model of its class and a benchmark for any more involved calculation
with 'correlated instantons'. Before studying more complicated ensembles, it
is important to understand the predictions of the random model  for physical
quantities. Our previous studies of the propagation of a
single quark \cite{Shuryak_Ver_il1} and quark-antiquark pairs (mesonic
correlators) \cite{Shuryak_Ver_il2} in such a vacuum have significantly
extended those reported in \cite{Shuryak_C}: the box volume was increased
by a factor 20, the correlation functions were traced down by a few orders
of magnitude, $etc.$. Furthermore, the results were shown to be in quantitative
 agreement with experimental data, which was not expected for this
simple model. Experimentally known correlators (and eventually, masses
and coupling constants of the mesons) are reproduced, in some cases literally
inside the uncertainties of the calculation!  What is equally exciting
is the fact that we also observe good agreement with the results  of
recent lattice calculations \cite{Negele_etal}, in spite of the quite different
approaches and approximations used.

As this paper is the first exploratory study of baryon correlators
in the instanton vacuum,  the use of the random model appears to be  well
justified. Its primary goal is to check whether this model leads at least
to some {\it baryonic bound states}, in the form of a resonances well
separated from a multi-particle continuum.

  Before we start, let us make a few  general
remarks about the baryonic channels.
It is not surprising that quark-quark forces are even
less understood than the quark-antiquark ones: ordinary (made of
light quarks) baryons are not simple two-body systems. The situation is
somewhat simpler if one of the quarks is 'sterile'
(very heavy) \cite{Shuryak_hl},
but unfortunately the experimental data on heavy-light baryons are still
rather incomplete.

A direct manifestation of the 't Hooft instanton-induced
interaction \cite{t_Hooft}
(which produces so spectacular effects for the scalar and
pseudoscalar mesonic correlators) is also expected in the quark-quark
interaction. In this case, it appears as a strong attraction in
the scalar diquark channel \cite{Betman_Laperashvili}.  As suggested in
\cite{Rosner_Shuryak}, this interaction can naturally explain all spin
splittings in the octet and the decuplet, describing the spectra at
least as good as traditional explanations using gluomagnetic forces
between 'constituent' quarks.  Thus, our second major question is
whether baryons belonging to the two lowest $SU(3)$ multiplets, the octet
and the decuplet, are {\it similar} (their differences  being due to
relatively small perturbative effects), or whether they are {\it completely
different}. In this context we not only want to consider the  $N-\Delta$ mass
splitting, but also make a comparison of the coupling constants and the
general shape of the correlation functions.

The paper is structured as follows: in section 2 we introduce 'diquark'
correlators, which are compared to the corresponding mesonic channels
and then used (in section 3) to derive properties of heavy-light
baryons. In sections 4 and 5 we deal with the nucleon and delta  correlators,
as representatives of the $SU(3)$ octet and decuplet baryons. A comparison
with QCD sum rules and lattice results,  as well as
a general discussion is contained in  sections 6 and 7. Conclusions
are summarized in section 8.

\renewcommand{\theequation}{2.\arabic{equation}}
\setcounter{equation}{0}
\vskip 1.5 cm
\centerline{\bf 2. Diquarks }
\vskip 0.5 cm

In this section we study correlation functions of two-quark currents.
As suggested in \cite{Shuryak_hl}, such currents can be thought of
as part of a baryon current in which one of the quarks is 'sterile', or
very heavy, so that it does not move at all.
We will come back to this interpretation in the next section.

The generic diquark current $j_\Gamma^a$ is defined by
\be
j_\Gamma^a = \epsilon^{abc} q^b_\alpha (C \Gamma)
  _{\alpha\beta} \tau q^c_\beta,
\ee
where $C$ is the charge conjugation matrix and $\Gamma$ is one of the
possible gamma matrix structures. Furthermore, $q^a_\alpha$ denotes an
isodoublet quark spinor with color index $a$ and spinor index $\alpha$.
The isospin matrix $\tau$ is given
by $\tau=\tau_2$ for an isosinglet diquark and by $\tau=\tau_2\vec\tau$
for an isovector diquark.  Note that the isospin  wavefunction is
determined by the symmetry of the Dirac matrix $C\Gamma$ and the
overall antisymmetry with respect to exchanging the two quarks.
In particular, $\Gamma=\{1\!\!1,\gamma_5,\gamma_\mu\gamma_5\}$
corresponds to isoscalar diquarks whereas $\Gamma=\{\gamma_\mu,
\sigma_{\mu\nu}\}$ gives isovector diquarks.

Note that the parity of the Dirac matrix $C\Gamma$ is opposite to that
of the matrix $\Gamma$: the internal parity of a $qq$ pair is
positive while the one of the $\bar qq$ system is negative.
For example, if $\Gamma=\gamma_5$, the diquark is a {\it scalar}.
This well known fact makes diquark notations
somewhat ambiguous, one can either (i) specify the real quantum numbers of the
diquarks, or (ii) indicate which gamma matrix was used in the current.
We  use the latter notation (e.g. call the $\Gamma=\gamma_5$ case
'the {\it pseudoscalar} correlator'), since in this notation the comparison
between diquark and meson correlators is more straightforward. We hope that
this will not confuse the reader.

We study diquark correlators defined by
\be
\Pi_\Gamma(x-y) = <j^a_\Gamma(x) \left( j^a_\Gamma(y)\right)^\dagger >,
\ee
where the hermitian conjugate current $(j_\Gamma^a)^\dagger$ is given by
\be
 (j^a_\Gamma)^\dagger = \epsilon^{abc}
  \bar q^b_\alpha (\Gamma C)_{\alpha\beta}\tau \bar q^c_\beta.
\ee
The correlator can be expressed in terms of the quark propagator
resulting in
\be
\label{diq_cor}
\Pi_\Gamma (x-y) = {\rm Tr}\left[ S^{ba}(x-y) \Gamma C S^{T\, ab}(x-y) C
\Gamma\right] - {\rm Tr} \left[ S^{aa}(x-y) \Gamma C S^{T\, bb}(x-y)
C \Gamma \right] \! ,
\ee
where the transpose is with respect to the Dirac indices.
At short distances we can evaluate the correlator using free
quark propagators. Deviations from the free quark propagation
have been studied using the operator product expansion (OPE), further
simplified by  the so called 'vacuum dominance' approximation \cite{SVZ}.
For baryonic currents it was implemented  \cite{Ioffe} in a very simple way:
the propagator is given by the sum of a free propagator and a quark
condensate. As discussed in more detail in \cite{Shuryak_hl}, this
approximation corresponds to a 'distance dependent constituent quark mass'
$m_Q=(\pi^2/3)\tau^2|\langle\bar qq\rangle |$ which is generated as the
quark propagates in the chirally asymmetric QCD vacuum.
The corresponding propagator reads\footnote{
In what follows we will always work in
euclidean space-time. Physically there is no preferred direction, so
we can always choose $x-y$ to point in the euclidean time direction.}
\be
\label{vac_dom}
iS(\tau) = -\frac 1{2\pi^2\tau^3}\gamma_0 - \frac 1{12} <\bar qq>.
\ee
As we have shown in the case of
the  mesonic correlators,  this approximation is quite useful in understanding
the experimental trend of the various correlation functions, although it is
far from being quantitatively correct. For the diquark correlators
introduced above we find
\setcounter{figure}{\value{equation}}
\renewcommand{\theequation}{2.6\alph{equation}}
\setcounter{equation}{0}
\be
\label{pis_vac_dom}
S:\quad  \frac {\Pi_S}{\Pi^0_S} &=& 1 - \asovbs,\\
\label{pip_vac_dom}
P:\quad  \frac {\Pi_P}{\Pi^0_P} &=& 1 + \asovbs,\\
\label{piv_vac_dom}
V:\quad  \frac {\Pi_V}{\Pi^0_V} &=& 1 + \frac{\pi^4}{18}
               |\langle\bar qq\rangle|^2 \tau^6 ,\\
A:\quad  \frac {\Pi_A}{\Pi^0_A} &=& 1 - \frac{\pi^4}{18}
               |\langle\bar qq\rangle|^2 \tau^6 ,\\
T:\quad  \frac {\Pi_T}{\Pi^0_V} &=& \;\;\frac{\pi^4}{6}
               |\langle\bar qq\rangle|^2 \tau^6 ,
\ee
\renewcommand{\theequation}{2.\arabic{equation}}
\addtocounter{figure}{1}
\setcounter{equation}{\value{figure}}
\noindent
where $\Gamma=\{S,P,V,A,T\}$ labels the {\it gamma matrices} in the
diquark channel and $\Pi^0_\Gamma$ denotes the free correlation function.
Note that these estimates for the short-distance behaviour of the
diquark correlation functions {\it exactly  coincide} with the
corresponding approximation for the mesonic correlator in the same
gamma matrix channel. This is a consequence
of the fact that in the vacuum dominance approximation the propagator
is diagonal in color space, which is not correct in general.

We now come to a physical interpretation of the diquark correlators.
Since diquarks are color {\it antitriplets}, they cannot
appear as asymptotic states. Their correlation functions are gauge-dependent
and unphysical, unless either (i) a gauge fixing is implied or (ii)
the additional color matrix  $U=P\exp(ig\int_C dx_\mu A^a_\mu t^a/2)$ is
included. If the path C is a segment of a straight line going in the time
direction, it corresponds to a static third quark: thus we naturally come to
light-heavy baryons, to be discussed in the next section.

  The specific feature of instanton models is that the matrix
$U\approx 1$, because heavy quarks interact very little with instantons.
We will therefore neglect the effects of $U$ in our following analysis.
More quantitatively, it was shown in \cite{Shuryak_C} that the gauge string
$U$ produces a  potential of the order of 50 MeV, which is well inside the
accuracy of the present calculations. Thus, {\it inside instanton-based
vacuum models} (which ignore confinement etc.), the diquark correlators
defined above are 'practically gauge invariant'\footnote{
The reader who does not like such an approximation, may
skip our qualitative discussion of diquarks and proceed directly
to our discussion of nucleon and delta correlators, which are
explicitly gauge invariant.}. In such a  spirit, one can define the
coupling of the currents introduced above to 'diquark states'
\setcounter{figure}{\value{equation}}
\renewcommand{\theequation}{2.7\alph{equation}}
\setcounter{equation}{0}
\be
  <0| j^a_{S,P}\,|\phi^b_{S,P}>\, &=& g_{S,P}
 \frac{\delta^{ab}}{\sqrt{3}}e^{-iqx}, \\
  <0| j^a_{V,A\,\mu}|\phi^b_{V,A}> &=& g_{V,A}
 \frac{\delta^{ab}}{\sqrt{3}}\epsilon_\mu e^{-iqx}, \\
  <0| j^a_{V,A\,\mu}|\phi^b_{S,P}> &=& f_{V,A}
 \frac{\delta^{ab}}{\sqrt{3}} q_\mu e^{-iqx} ,
\ee
\renewcommand{\theequation}{2.\arabic{equation}}
\addtocounter{figure}{1}
\setcounter{equation}{\value{figure}}
\noindent
where $|\phi^a_{\Gamma}>$ denotes a diquark state with momentum $q_\mu$
and color index $a$
that has the quantum numbers of the corresponding current and
$\epsilon_\mu$ is the polarization of a vector diquark.
Proceeding this way we effectively treat diquarks just like mesons.
In particular, we also allow for mixing of vector and scalar
diquarks in the vector channel. This phenomenon is analogous to
the well known $\pi-a_1$ mixing which is observed in the
axial vector channel.

In the following we want to describe the diquark correlation
functions in terms of the physical intermediate states
which can contribute. For this reason we represent the
euclidean Fourier transform of the correlation function by
a dispersion integral
\be
\label{disp_rel}
  \int d^4x\, e^{iqx}\Pi_\Gamma(x) = \frac{1}{\pi}\int_{0}^{\infty}
    ds\frac{{\rm Im}\Pi_\Gamma(s)}{q^2+s}\, ,
\ee
where $\rho_\Gamma(s)\equiv\frac{1}{\pi}{\rm Im}\Pi_\Gamma(s)$ is the
spectral density associated with the corresponding correlation function.
The spectral density can be expressed as
\be
\label{spec_dens}
 \rho_\Gamma(s=k^2) = (2\pi)^3 \sum_{n} \delta^4(k-q_n)
     <0|j_\Gamma(0)|n><n|j_\Gamma^\dagger(0)|0>\, ,
\ee
where $|n>$ is a physical intermediate state with momentum $q_n$.

Similar to the procedure
in the meson sector we approximate the spectral function
by the contribution of one or two diquark resonances and a free
quark continuum starting at a threshold $s_0$. This means
that for long distances the correlator is dominated by the
contribution of the lowest physical state whereas for short
distances it can be described in terms of the propagation
of free quarks. In the scalar and pseudoscalar channels we get
\be
\label{rho_sp}
 \rho_{S,P}(s) = g_{S,P}^2 \delta (s-m_{S,P}^2)
    +\frac 3{4\pi^2} s\Theta(s-s_0) ,
\ee
where the first term is the resonance contribution and the
second term represents the continuum. On purely dimensional
grounds we have $\rho_{S,P}(s)\sim s$ for large invariant masses.
The corresponding coefficient can be calculated from the
discontinuity of the free quark loop diagram appearing in
eq.(\ref{diq_cor}).

Inserting the model eq.(\ref{rho_sp}) for the spectral function
in the dispersion relation (\ref{disp_rel}) and transforming back
to coordinate space the diquark correlator is given by
\be
\label{sp_cor}
 \Pi_{S,P}(\tau) = g_{S,P}^2 D(m_{S,P},\tau) +
    \frac 3{4\pi^2}\int_{s_0}^\infty ds \,s D(\sqrt{s},\tau)\, ,
\ee
where $D(m,\tau)=m/(4\pi^2\tau) K_1(m\tau)$ is the euclidean space
propagator for a massive scalar particle. Proceeding in the same
way in the vector channels, we find
\be
\label{va_cor}
 \Pi_{V,A}(\tau) = 3g_{V,A}^2 D(m_{V,A},\tau)-
           m_{S,P}^2 f_{V,A}^2 D(m_{S,P},\tau) +
    \frac 3{2\pi^2}\int_{s_0}^\infty ds \,s D(\sqrt{s},\tau)\, ,
\ee
In the tensor channel, finally, there is no continuum contribution.
The dirac tensor $\sigma_{\mu\nu}$ has components $\sigma_{0i}$
which transform like a three-vector and  $\epsilon_
{ijk}\sigma_{jk}$ which transform like a pseudovector. This means
that vector as well as axial vector diquark states can contribute.
Since the contributions
from these two states are hard to distinguish in practice we
describe the tensor diquark correlation function using a single
diquark resonance with mass $m_T$ and coupling $g_T$
\be
\label{t_cor}
 \Pi_{T}(\tau) = g_T^2 D(m_T,\tau) .
\ee
We have extracted the diquark masses and coupling constants by
fitting the parametrizations eq.(\ref{sp_cor}-\ref{t_cor}) to
the diquark correlation functions measured in the random instanton
liquid. The correlation functions together with the fits are
shown in figure 1 and the results of the fit are summarized in
table 1.

Looking at the various diquark correlation functions one immediately
realizes that there are very pronounced qualitative differences
between the different channels. In particular, the scalar case ($\Gamma=1$)
has very repulsive interaction and shows no indications for a 'diquark
resonance' whereas the pseudoscalar one ($\Gamma=\gamma_5$)
 shows a huge enhancement\footnote{Which is still not as strong as for
the pseudoscalar mesons, of course, which have to be massless due to Goldstone
theorem. However, it is quite comparable with the effect observed for scalar
isoscalar 'sigma' meson, which also has a comparable mass.}, or strong
attractive interaction. Determining the parameters of this
resonance we find
\be
  m_P = 420\pm 30\,{\rm MeV},\hspace{1cm}
  g_P = 0.23\pm 0.01 {\rm GeV}^2 .
\ee

The dashed curve in figure 1 shows the result of the 'vacuum dominance'
approximation,  see \cite{Shuryak_Ver_il2}.
As for the mesonic channels, it reproduces the qualitative differences
between the scalar and pseudoscalar channels but fails to give
a quantitative description a distances larger than 0.3 fm.

In the ($\Gamma=\gamma_\mu$)
vector diquark channel the correlation function remains
close to the perturbative one out to fairly large distances. Again,
the situation is similar to the superduality phenomenon observed
in the rho meson channel. The resonance parameters in the diquark
channel are
\be
  m_V = 940\pm 20\,{\rm MeV},\hspace{1cm}
  g_V = 0.24\pm 0.01 {\rm GeV}^2 .
\ee
In the ($\Gamma=\gamma_\mu\gamma_5$) axialvector diquark channel,
we can have mixing with the light pseudoscalar diquark. Since the
contribution from this state alone can already explain the measured
correlation function, we have no evidence for an axialvector
'diquark state' in the RILM. In the tensor ($\Gamma=\sigma_{\mu\nu}$)
diquark channel we have fitted the correlator with a single resonance
at $m_T=570\pm 20$ MeV, which is intermediate between the light
pseudoscalar and heavy vector diquark resonances.

Of great interest for hadronic models which invoke diquark clustering
in order to explain the observed spin splitting between the octet and
the decuplet is the mass difference $m_V-m_P\simeq 520$ MeV. In
a simple quark-diquark model \cite{Rosner_Shuryak,Lichtenberg_etal} we
have $m_\Delta-m_N =\frac 12 (m_V-m_P)$. Comparing with the experimental
splitting $m_\Delta-m_N=293$ MeV we do in fact get a very good agreement.
We will further comment on the quark-diquark picture of the baryons in
our discussion of the nucleon and delta correlation functions.

\vskip 1.5cm
\renewcommand{\theequation}{3.\arabic{equation}}
\setcounter{equation}{0}
\centerline{\bf 3. Heavy-light baryons }
\vskip 0.5 cm
   Adding a heavy quark $Q$ to the diquark currents considered in the
last section we can construct currents carrying the quantum numbers
of heavy baryons. Correlation functions of these currents were first
considered in \cite{Shuryak_hl} in the framework of the QCD sum rules.
Recently, further analysis along these lines  was performed in
\cite{Grosin_etal,Bagan_etal}, where somewhat different (non-relativistic)
currents have been used.

The heavy baryon  currents have the general
structure $J_{B_Q\,\alpha}=j_\Gamma(\Gamma^\prime Q)_\alpha$ where
$j_\Gamma$ is a diquark current, $Q$ is a heavy quark spinor with
dirac index $\alpha$ and $\Gamma^\prime$ is a dirac matrix.
Out of the five possible currents,
three couple to the $\Lambda_Q$ while the two remaining ones have
the quantum numbers of the $\Sigma_Q$:
\setcounter{figure}{\value{equation}}
\renewcommand{\theequation}{3.1\alph{equation}}
\setcounter{equation}{0}
\be
 & J_{\Lambda_Q}^1 = \epsilon^{abc} (u^a Cd^b)\gamma_5 Q^c
  \hspace{0.5cm}
  J_{\Lambda_Q}^2 = \epsilon^{abc} (u^a C\gamma_5d^b) Q^c
 & \nonumber \\
\label{lam_cur}
 & J_{\Lambda_Q}^3 = \epsilon^{abc} (u^a C\gamma_\mu\gamma_5d^b)
      \gamma^\mu Q^c ,& \\
\label{sig_cur}
 &  J_{\Sigma_Q}^1 = \epsilon^{abc} (u^a C\gamma_\mu u^b)
     \gamma_5\gamma^\mu Q^c
  \hspace{0.5cm}
  J_{\Sigma_Q}^2 = \epsilon^{abc} (u^a C\sigma_{\mu\nu}u^b)
  \gamma_5\sigma^{\mu\nu} Q^c\, . &
\ee
\renewcommand{\theequation}{3.\arabic{equation}}
\addtocounter{figure}{1}
\setcounter{equation}{\value{figure}}
\noindent
In the following we will consider correlation functions
constructed from these currents
\be
\label{heavy_light_cor}
  \Pi_{\alpha\beta}(x-y) = i< J_{B_Q\,\alpha}(x)
    \bar J_{B_Q\,\beta}(y) >.
\ee
It is instructive to consider the limit in which the mass
of the heavy quark is taken to infinity.
In this case the correlation function simplifies to
\be
\label{eff_cor}
  \Pi_{\alpha\beta}(x-y) = (\Gamma^\prime iS_Q(x-y)
     \overline\Gamma^\prime)_{\alpha\beta}  \, \Pi_\Gamma(x-y) ,
\ee
where $\overline\Gamma^\prime=\gamma_0\Gamma^{\prime\dagger}\gamma_0$
and
\be
iS_Q(x-y) = \frac{1+\gamma_0}{2}\Theta(\tau)\delta^3(\vec x-\vec y)
\left ( \frac M{2\pi\tau} \right )^{\frac 32} e^{-M\tau}
\ee
 denotes the heavy quark propagator in the nonrelativistic (static) limit.
Using this approximation
the spin structure of the correlation function becomes trivial
and all the dynamical information is contained in the diquark
correlator $\Pi_\Gamma(x-y)$.

As in the last section we want to describe the Euclidean correlation function
(\ref{heavy_light_cor}) in terms of physical intermediate states. This
is achieved by means of the spectral decomposition of the correlator
\be
\Pi_{\alpha\beta}(\tau) =
\int_0^\infty ds \rho_{\alpha\beta}(s) D(\sqrt s, \tau)
\ee
Again we approximate the spectral function by the sum of a resonance
and a continuum contribution, which in this case results in
\be
\rho_{\alpha\beta}(q) &=& \delta (q^2-M^2_{B_Q})
2M_Q\lambda_Q^2 \sum_s Q_\alpha^{(s)}
\overline{Q}_\alpha^{(s)} \\
   & & \; \;+\; \frac 1{10\pi^4}
(s-s_0)^{5/2}\theta(s-s_0)( (\gamma\cdot
q)_{\alpha\beta} + M_Q\delta_{\alpha\beta}) \nonumber .
\ee
The coupling constant $f_B$ is defined by
\be
<0|J_{B_Q \alpha}|B_Q> =  f_B \sqrt{M_Q} B_\alpha^{(s)},
\ee
where $B_\alpha^{(s)}$ is a Dirac spinor with spin $s$.
The numerical factor multiplying the continuum contribution follows from
asymptotic freedom.

It is convenient to measure the energies with respect to the heavy quark mass
$M_Q$. When we write $s = (M_Q + \omega)^2$, the non-relativistic heavy
quark propagator factorizes from the spectral function, and the Euclidean
diquark correlator is given by
\be
\label{non_rel_fit}
 \Pi_\Gamma(\tau) = \frac{f^2_{B_Q}}{2}\exp (-\epsilon_{B_Q}\tau)
        + \frac{1}{20\pi^4}\int_{\omega_0}^\infty
        d\omega \,\omega^5 \exp(-\omega\tau)\, ,
\ee
where $\epsilon_{B_Q}$ is the heavy baryon energy (minus the
rest mass of the heavy quark).
As above we have introduced a continuum
integral starting at a threshold energy $\omega_0$.

The heavy baryon parameters extracted from this non-relativistic
interpretation of the diquark correlation functions are given
in table 2, together with the predictions from QCD sum rules.

 The $\Lambda_Q$ results were obtained from the
pseudoscalar diquark channel, since this current has a simple
non-relativistic limit. We find
\be
 \epsilon_{\Lambda_Q}=760\pm 30\,{\rm MeV} \hspace{1cm}
 f_{\Lambda_Q} = 0.052\pm 0.005\,{\rm GeV}^3\, .
\ee
The parameters of the $\Sigma_Q$ baryon can be determined from
the $\Gamma=\gamma_\mu$ or $\Gamma=\sigma_{\mu\nu}$ diquark channels.
Using the $\Gamma=\gamma_\mu$ diquark channel we get
\be
 \epsilon^V_{\Sigma_Q}= 892\pm 30 \,{\rm MeV}, \hspace{1cm}
 f^V_{\Sigma_Q} =  0.011 \pm 0.002 \,{\rm GeV}^3\, .
\ee
while the fit to the tensor channel gives a somewhat larger mass\footnote{
Naturally, the two coupling constants
$f^V_{\Sigma_Q},f^T_{\Sigma_Q}$ have different definitions and should
not be equal.}
\be
 \epsilon^T_{\Sigma_Q}= 1006\pm 30 \,{\rm MeV}, \hspace{1cm}
  f^T_{\Sigma_Q} =  0.044 \pm 0.008 \,{\rm GeV}^3\, .
\ee
Of particular interest is the  $\Sigma_Q-\Lambda_Q$ mass splitting
which we find to be $m_{\Sigma_Q}-m_{\Lambda_Q}=135$ MeV or 246 MeV,
depending on the current used in the fit. This
number should be compared with $m_{\Sigma_c}-m_{\Lambda_c}=
170$ MeV for charmed baryons. Unfortunately, the corresponding
value for bottom baryons is not known experimentally.

   It is instructive to compare these results with those that have
been obtained from QCD sum  rules. In the original paper \cite{Shuryak_hl}
the predictions are\footnote{There was a misprint in the paper:
The value of $f_{\Lambda_Q}$ is one order of magnitude larger than
indicated. The published version corresponds to the r.h.s. of the sum rules
 100 times smaller than the l.h.s.}
$
 \epsilon_{\Lambda_Q}=700\,{\rm MeV}$ and $
 f_{\Lambda_Q} = 0.02\,{\rm GeV}^3
$
together with a mass splitting $m_{\Sigma_Q}-m_{\Lambda_Q}=400\pm 250$ MeV.
On the other hand, the values obtained in \cite{Grosin_etal} are
$
 \epsilon_{\Lambda_Q}=780\,{\rm MeV}$ and $
 f_{\Lambda_Q} = 0.022\pm 0.005\,{\rm GeV}^3
% \epsilon_{\Sigma_Q}=990\,{\rm MeV} \hspace{1cm}
% \hat f_{\Lambda_Q} = 0.029-0.041\,{\rm GeV}^3\, .
$
which is very similar, but there is {\it no} $\Sigma_Q-\Lambda_Q$ splitting!
Let us make a few remarks concerning these results:

(i) The two OPE-based papers give  similar predictions for the
mass and the coupling constant of the $\Lambda_Q$ when the pseudoscalar
current is used. While the predicted mass agrees with our fit,
the coupling constant does not. This is not surprising:  the large
instanton-induced attractive interaction which leads to the strong
enhancement in the $\gamma_5$ channel is {\it not} reproduced by the OPE.

(ii) Extracting the $\Sigma_Q$ parameters
and the $\Lambda_Q-\Sigma_Q$ splitting from the  OPE is very subtle: in
fact the 'vacuum dominance' approximation (inclusion of the quark
condensate term in the propagator) {\it does not lead to an
unambiguous result}.

   Indeed, compare for example the $\gamma_\mu$ and
$\gamma_\mu \gamma_5$ correlators in order to determine the
$\Sigma_Q-\Lambda_Q$ splitting. In this case the corrections up to order
$\langle\bar qq\rangle^2$ read $1\pm\langle\bar qq\rangle^2
\tau^6\pi^4/18$, respectively, so that one may think there is
a big splitting. However, if one considers the non-relativistic limit of the
currents (see ref. \cite{Grosin_etal}),
using the  $\vec \gamma$ diquark  for the nonrelativistic
$\Sigma_Q$ current and $\gamma_0 \gamma_5$ for the $\Lambda_Q$, one
finds {\it identical} OPE expressions $1+\langle\bar qq\rangle^2
\tau^6\pi^4/36$ in both cases, leading to {\it zero}\footnote{The
contribution of the gluon condensate also turns out to be
identical.} mass splitting!

(iii) Our fits for the $\Sigma_Q$ mass in the $V$ and $T$ channels are
different, but not very much so. A $\Sigma_Q-\Lambda_Q$ splitting on the order
of 200 MeV is definitely predicted. What is maybe even more important is that
we see a qualitative difference between the $\Sigma_Q$ and the $\Lambda_Q$ :
both $V$ and $T$ correlators are comparable to
the perturbative ones at distances about 1 fm, while the correlator in the
$\gamma_5$ channel is several times larger.

\vskip 1.5 cm
\renewcommand{\theequation}{4.\arabic{equation}}
\setcounter{equation}{0}
\centerline{\bf 4. Nucleon correlation functions }
\vskip 0.5 cm
After studying the quark-quark interaction we now proceed to
correlation functions of baryonic (three quark) currents.
The Euclidean correlator of two spin 1/2 nucleon currents is defined  as
\be
\label{nuc_cor}
\Pi_{\alpha\beta}(x-y) = i< J^N_\alpha(x) \bar J^N_\beta(y)>,
\ee
where $\alpha,\beta$ are the spinor indices of the nucleon currents.
Using Lorentz and parity invariance one can show
that the correlator can be decomposed in terms of only two independent
Dirac structures
\be
\label{nuc_cor_dec}
\Pi_{\alpha\beta}(x-y) = \Pi_1((x-y)^2) (\gamma\cdot (x-y))_{\alpha\beta} +
\Pi_2((x-y)^2) \delta_{\alpha\beta}.
\ee
Again we will consider the correlation function in the euclidean time
direction. Only two independent correlators remain for a given current,
namely ${\rm Tr}( \Pi (\tau) \gamma_0)$ and ${\rm Tr} (\Pi (\tau))$.

Due to isospin symmetry it is sufficient to consider only one charge
state of the nucleon. A proton current is constructed by coupling
a $d$-quark to a  $uu$-diquark. The diquark has the structure
$\epsilon_{abc} u_b C\Gamma u_c$ which requires that the matrix
$C\Gamma$ is symmetric. This condition is satisfied for the $V$ and
$T$ gamma matrix structures. Only two possible currents that have positive
parity and spin $1/2$ can be constructed \cite{Ioffe}. These so called
{\it Ioffe currents} are given by
\setcounter{figure}{\value{equation}}
\renewcommand{\theequation}{4.3\alph{equation}}
\setcounter{equation}{0}
\be
\eta_1 &=& \epsilon_{abc} (u^a C\gamma_\mu u^b) \gamma_5 \gamma_\mu d^c, \\
\eta_2 &=& \epsilon_{abc} (u^a C\sigma_{\mu\nu} u^b) \gamma_5 \sigma_{\mu\nu}
 d^c .
\ee
\renewcommand{\theequation}{1.\arabic{equation}}
\addtocounter{figure}{1}
\setcounter{equation}{\value{figure}}
\noindent
In total, there exist six different nucleon correlators: the diagonal
$\eta_1\bar\eta_1,\eta_2\bar\eta_2$ and off-diagonal $\eta_1\bar\eta_2$
 correlators, each contracted with either the identity or
$\gamma_0$.
%
%The total number of currents can also be understood in a different way.
%The spinors can be decomposed into left-handed and right-handed pieces,
%$u = u_R+u_L$ and $d = d_R + d_L$. This allows us to construct 4
%possible currents
%\setcounter{figure}{\value{equation}}
%\renewcommand{\theequation}{4.4\alph{equation}}
%\setcounter{equation}{0}
%\be
%J_{RRR} &=& (u_R C d_R) u_R,\\
%J_{RRL} &=& (u_R C d_R) u_L,\\
%J_{LLR} &=& (u_L C d_L) u_R,\\
%J_{LLL} &=& (u_L C d_L) u_L,
%\ee
%\renewcommand{\theequation}{1.\arabic{equation}}
%\addtocounter{figure}{1}
%\setcounter{equation}{\value{figure}}
%
%\noindent
%where the Dirac and color indices have been suppressed.
%The Ioffe currents are simple linear combinations of these
%currents\footnote{Note that a factor 2 is missing in eq. (62)
%of ref. \cite{Ioffe}}:
%$\eta_1 = 4(J_{RRL} -J_{LLR})$ and $\eta_2 = 8(J_{LLL} -J_{RRR})$
%Noting that
%interchanging all left-handed and all right-handed spinors does not
%affect the correlator, we again find six different correlators.
They are defined as
\setcounter{figure}{\value{equation}}
\renewcommand{\theequation}{4.4\alph{equation}}
\setcounter{equation}{0}
\be
\Pi_1^N &=& \frac 1{4} <{\rm Tr} (\eta_1\bar\eta_1)>,\\
\Pi_2^N &=& \frac 1{4} <{\rm Tr} (\gamma_0 \eta_1\bar\eta_1)>,\\
\Pi_3^N &=& \frac 1{4} <{\rm Tr} (\eta_2\bar\eta_2)>,\\
\Pi_4^N &=& \frac 1{4} <{\rm Tr} (\gamma_0\eta_2\bar\eta_2)>,\\
\Pi_5^N &=& \frac 1{4} <{\rm Tr} (\eta_1\bar\eta_2)>,\\
\Pi_6^N &=& \frac 1{4} <{\rm Tr} (\gamma_0 \eta_1\bar\eta_2)>,
\ee
\renewcommand{\theequation}{4.\arabic{equation}}
\addtocounter{figure}{1}
\setcounter{equation}{\value{figure}}
\noindent
Again we can discuss the qualitative features of these correlation functions
using the vacuum dominance model (\ref{vac_dom}). The result is
\setcounter{figure}{\value{equation}}
\renewcommand{\theequation}{4.5\alph{equation}}
\setcounter{equation}{0}
\be
\frac{\Pi_1^N}{\Pi_2^{N0}} &=& \;\;\,\frac{\pi^2}{12}
            |\langle\bar qq\rangle| \tau^3 + \aqovbq,\\
\frac{\Pi_2^N}{\Pi_2^{N0}} &=& 1 + \frac{\pi^4}{72}
            |\langle\bar qq\rangle|^2 \tau^6 ,\\
\frac{\Pi_3^N}{\Pi_4^{N0}} &=& \;\;\,\frac{\pi^6}{216}
            |\langle\bar qq\rangle|^3 \tau^9,\\
\frac{\Pi_4^N}{\Pi_4^{N0}} &=& 1 ,\\
\frac{\Pi_5^N}{\Pi_2^{N0}} &=& \;\;\,\frac{\pi^2}{2}
            |\langle\bar qq\rangle| \tau^3,\\
\frac{\Pi_6^N}{\Pi_2^{N0}} &=& \;\;\,\frac{\pi^4}{12}
            |\langle\bar qq\rangle|^2 \tau^6.
\ee
\renewcommand{\theequation}{4.\arabic{equation}}
\addtocounter{figure}{1}
\setcounter{equation}{\value{figure}}
\noindent
In order to compare these functions with experimental information
we have to analyze the correlation functions in terms of
physical intermediate states. The nucleon coupling to the
two Ioffe currents is given by
\be
\label{nuc_coupling}
   <0|\eta_{1,2}(0) |N(q,s)> &=& \lambda^N_{1,2}
     \sqrt{\frac{2m_N}{(2\pi)^3}} u(q,s)e^{-iqx},
\ee
where $|N(q,s)>$ denotes a nucleon state with momentum $q$ and
spin $s$, $u(q,s)$ is the corresponding free fermion spinor
(normalized as in \cite{BJORKEN-DRELL-1965}) and
$\lambda_{1,2}^N$ is a coupling constant. The first two
correlation functions introduced above are related to the
first Ioffe current. Using the simple nucleon resonance plus
continuum model introduced in section 2 we have
\setcounter{figure}{\value{equation}}
\renewcommand{\theequation}{4.7\alph{equation}}
\setcounter{equation}{0}
\be
 \Pi_1^N(\tau) &=& \;\;\;\left(\lambda^N_1\right)^2 m_N D(m_N,\tau )
   + \frac{1}{4\pi^2}|\langle\bar qq\rangle|
      \int_{s_0}^{\infty} ds\, s D(\sqrt s,\tau) ,\\[0.1cm]
 \Pi_2^N(\tau) &=& -\left(\lambda^N_1\right)^2 D^\prime (m_N,\tau )
   - \frac{1}{2^{6}\pi^4}
      \int_{s_0}^{\infty} ds\, s^2 D^\prime (\sqrt s,\tau),
\ee
\renewcommand{\theequation}{4.\arabic{equation}}
\addtocounter{figure}{1}
\setcounter{equation}{\value{figure}}
\noindent
where $D(m,\tau)$ is the euclidean space propagator of a scalar
particle of mass $m$ and $D^\prime(m,\tau)=d/(d\tau) D(m,\tau)$.
For the second correlation function $\Pi_2^N$ the continuum corresponds
to the propagation of three free quarks and can be calculated from the
discontinuity of the quark loop diagram contributing to the
correlation function of the two Ioffe currents.

In the case of $\Pi_1^N$ there is no such contribution. In
order to improve the description at short distances we have added
the discontinuity of the diagram in which one
of the quarks interacts with the quark condensate. In the language
of the operator product expansion this means that we have calculated
the spectral function from the imaginary parts of the diagrams
which determine the coefficients of the unit and $\langle\bar qq
\rangle$-operators. In a more physical picture one can check that,
at least in an average sense, the continuum part of $\Pi_1^N$
represents the contribution of the $\pi\Delta$-continuum and
higher nucleon resonances.

The correlation functions $\Pi^N_{3,4}$ are related to the
second Ioffe current. Proceeding as above they can be written as
\setcounter{figure}{\value{equation}}
\renewcommand{\theequation}{4.8\alph{equation}}
\setcounter{equation}{0}
\be
 \Pi_3^N(\tau) &=& \;\;\;\left(\lambda^N_2\right)^2 m_N D(m_N,\tau ),
                        \\[0.3cm]
 \Pi_4^N(\tau) &=& -\left(\lambda^N_2\right)^2 D^\prime (m_N,\tau )
   - \frac{3}{2^{5}\pi^4}
      \int_{s_0}^{\infty} ds\, s^2 D^\prime (\sqrt s,\tau) .
\ee
\renewcommand{\theequation}{4.\arabic{equation}}
\addtocounter{figure}{1}
\setcounter{equation}{\value{figure}}
\noindent
The last two nucleon functions are related to off-diagonal
correlators between the two Ioffe currents. They are given by
\setcounter{figure}{\value{equation}}
\renewcommand{\theequation}{4.9\alph{equation}}
\setcounter{equation}{0}
\be
 \Pi_5^N(\tau) &=& \;\;\; \lambda^N_1\lambda^N_2
  m_N D(m_N,\tau) + \frac{3}{2\pi^2}|\langle\bar qq\rangle|
      \int_{s_0}^{\infty} ds\, s D(\sqrt s,\tau), \\[0.1cm]
 \Pi_6^N(\tau) &=& -\lambda^N_1\lambda^N_2
          D^\prime (m_N,\tau).
\ee
\renewcommand{\theequation}{4.\arabic{equation}}
\addtocounter{figure}{1}
\setcounter{equation}{\value{figure}}
The measured nucleon correlation functions together with the
physical parametrizations introduced above are shown in figure 3.
The dashed lines show the result of the vacuum dominance
approximation. Nucleon parameters determined from fitting the
correlation functions in the RILM are given in table 3.

The most important feature of the nucleon correlation functions
is the strong enhancement over the free correlator which can
be seen in $\Pi_2^N$ and $\Pi_4^N$. This behavior is reminiscent of the
pseudoscalar diquark channel which shows a  similar strong attractive
interaction. Somewhat surprisingly, {\it for the nucleon case} this feature
is semi-quantitatively reproduced by the vacuum dominance
approximation.

 This enhancement in the calculated correlators is nicely described by a
nucleon contribution to the correlation function. Fitting
the nucleon parameters we get
\be
  m_N=960\pm 30\,{\rm MeV}, \hspace{0.6cm}
  \lambda_1^N = 0.031\pm 0.001 \,{\rm GeV}^3, \hspace{0.6cm}
  \lambda_2^N = 0.080\pm 0.004 \,{\rm GeV}^3 ,
\ee
with a surprisingly accurate mass value. What is even more important,
the simple 'nucleon pole plus continuum' model gives
a very good simultaneous description for {\it the complete set} of
correlation functions.

This agreement is particularly good for correlators
involving  the {\it first} Ioffe current, while it is
somewhat worse for the other ones. Note that the correlation functions
$\Pi_3^N$  and $\Pi_6^N$ (which show deviations from the
simple 'nucleon pole plus continuum' model) involve
at least two quarks which have to flip their chirality.
It might be that the RILM has problems reproducing those amplitudes. An
alternative explanation (suggested by Ioffe in his original QCD sum
rule analysis) might be that the second Ioffe current couples more
strongly to some excited states of the nucleon.

It is important to note that the simple random instanton model,
even without confinement, {\it does create a nucleon bound state}. The
presence of a bound state is indicated by the fact that we can
describe the correlation functions in terms of a single pole,
{\it clearly separated} from the free quark continuum starting at
$E_0\simeq 2$ GeV.

In order to strengthen that point we have also considered other
models for the nucleon correlation functions.
The simplest one is a 'constituent quark model'. One may try to
 describe the correlator in terms of
the propagation of three uncorrelated 'constituent' quarks.
For their masses we take the coordinate
dependent effective mass $m_Q=(\pi^2/3)\tau^2|\langle\bar
qq \rangle|^2$ at short distances and $m_Q=300$ MeV at
large distances. The result is shown in figure 5 for the case
of $\Pi_2^N$. The conclusion is obvious: the
constituent model clearly can not explain the shape and magnitude
of the correlation function.

A somewhat less trivial model for the correlator is the
quark-diquark description. For this purpose we have Fierz
rearranged $\eta_1$ in such a way as to make the scalar
and vector diquark content explicit \cite{Ioffe}. We can
now use the {\it measured diquark correlators} which show a strong
enhancement in the scalar diquark channel, and describe the
propagation of the third quark using the constituent quark
propagator introduced above.
Again, the result is presented in figure 5. The corresponding
curve is much closer to the data than the pure constituent
model, but it still fails to reproduce our numerical results.

We conclude form these comparisons that the RILM not only
produces substantial attraction in the scalar diquark channel,
but also describes the nucleon as a true three quark bound
state. While the first feature is essentially a consequence of the
lowest order t'Hooft interaction, the presence of a nucleon pole
is a very non trivial consequence of the RILM.

\vskip 1.5 cm
\renewcommand{\theequation}{5.\arabic{equation}}
\setcounter{equation}{0}
\centerline{\bf 5. Delta correlation functions }
\vskip 0.5 cm
In the case of the $\Delta-$current, isospin symmetry allows us to
restrict ourselves to the $\Delta^{++}$. In this case all quarks have
the same flavors and with the help of a Fierz transformation one can
show \cite{Ioffe} that there exists only one independent
current which is given by
\be
J^\Delta_\mu =  \epsilon_{abc} (u^a C\gamma_\mu u^b)  u^c.
\ee
However, the spin structure of the correlator
\be
\Pi^\Delta_{\mu\nu;\alpha\beta}(x-y) &=& i<J^\Delta_{\mu\alpha}(x) \bar
  J^\Delta_{\nu\beta}(y)>
\ee
is much richer. As before the correlator is related to the spectral
function by
\be
\Pi^\Delta_{\mu\nu;\alpha\beta}(x) = \int ds \rho_{\mu\nu;\alpha\beta}
D(\sqrt s, x) .
\ee
For fixed values of $\mu$ and $\nu$ we again have a two-spinor
spectral function which, as seen in section 4, reduces to two invariant
structures which can be projected out by either contracting with the identity
or $\gamma_0$. The $\mu\nu-$indices transform as a Lorentz vector. Five
different vector structures are possible
\be
\delta_{\mu\nu},\quad \gamma_\mu \gamma_\nu, \quad q_\mu q_\nu, \quad
\gamma_\mu q_\nu, \quad \gamma_\nu q_\mu,
\ee
which adds up to ten different possibilities for the correlator. However,
the delta current satisfies the Rarita Schwinger constraint
$\gamma^\mu J_\mu^\Delta=0$ which allows us to express
the spectral function $\rho_{\mu\nu;\alpha\beta}$
using only four independent form factors
\be
\rho_{\mu\nu;\alpha\beta} &=& a_1(q^2)\,(g_{\mu\nu} \hat q - 6
         \frac{q_\mu q_\nu}{q^2}\hat q
     + \gamma_\mu q_\nu + \gamma_\nu q_\mu )\nonumber \\
 && + \;a_2(q^2)\, (\gamma_\mu\gamma_\nu \hat q - 16
      \frac{q_\mu q_\nu}{q^2}\hat q
      + 2 \gamma_\mu q_\nu + 4 \gamma_\nu q_\mu )\nonumber \\
 && + \;b_1(q^2)\,(-4 g_{\mu\nu} + \gamma_\mu\gamma_\nu )\nonumber \\
 && + \;b_2(q^2)\,( g_{\mu\nu}+ 2\frac{q_\mu q_\nu}{q^2}
     -\frac{\gamma_\mu q_\nu}{q^2}\hat q
     + \frac{\gamma_\nu q_\mu}{q^2} \hat q ).
\ee
The above $\Delta-$current does not satisfy the second Rarita-Schwinger
constraint $q^\mu J_\mu^\Delta = 0$. This condition would lead to two further
relations between the form factors
\setcounter{figure}{\value{equation}}
\renewcommand{\theequation}{5.6\alph{equation}}
\setcounter{equation}{0}
\be
a_1(q^2) &=& -3 a_2(q^2),\\
b_1(q^2) &=& b_2(q^2).
\ee
\renewcommand{\theequation}{5.\arabic{equation}}
\addtocounter{figure}{1}
\setcounter{equation}{\value{figure}}
\noindent
However, as will be shown below, the numerical values for these coefficients
do not differ much from this relation. Consequently, the admixture of spin
1/2 states to the spectral function will be small.

We take the correlator in the euclidean time direction and consider the
following invariant functions
\setcounter{figure}{\value{equation}}
\renewcommand{\theequation}{5.7\alph{equation}}
\setcounter{equation}{0}
\be
\Pi_1^\Delta &=& \frac 14{\rm Tr}( \Pi^\Delta_{\mu\nu}\delta_{\mu\nu}),\\
\Pi_2^\Delta &=& \frac 14{\rm Tr}( \Pi^\Delta_{\mu\nu}\delta_{\mu\nu}
                     \gamma_0),\\
\Pi_3^\Delta &=& \frac 14{\rm Tr}( \Pi^\Delta_{\mu\nu} \delta_{\mu 0}
                      \delta_{\nu 0} ),\\
\Pi_4^\Delta &=& \frac 14{\rm Tr}( \Pi^\Delta_{\mu\nu} \delta_{\mu 0}
                      \delta_{\nu 0} \gamma_0 ).
\ee
\renewcommand{\theequation}{5.\arabic{equation}}
\addtocounter{figure}{1}
\setcounter{equation}{\value{figure}}

As is section 3, we consider the correlators using the vacuum dominance
approximation. For the four  different structures we find
\setcounter{figure}{\value{equation}}
\renewcommand{\theequation}{5.8\alph{equation}}
\setcounter{equation}{0}
\be
 \frac{\Pi_1^\Delta} {\Pi_2^{\Delta 0}} &=& \;\;\, \frac{\pi^2}{3}
       |\langle\bar qq\rangle| \tau^3
      + \frac{\pi^6}{108} |\langle\bar qq\rangle|^3 \tau^9   ,\\
 \frac{\Pi_2^\Delta} {\Pi_2^{\Delta 0}} &=& 1 + \frac{\pi^4}{12}
       |\langle\bar qq\rangle|^2 \tau^6 ,\\
 \frac{\Pi_3^\Delta} {\Pi_4^{\Delta 0}} &=& \;\;\,\aovb  - \aqovbq,\\
 \frac{\Pi_4^\Delta}{\Pi_4^{\Delta 0}}  &=& 1 - \asovbs
\ee
\renewcommand{\theequation}{5.\arabic{equation}}
\addtocounter{figure}{1}
\setcounter{equation}{\value{figure}}
\noindent
As usual  we represent the spectral function for the delta correlator
by a resonance contribution and a continuum of states starting at an
invariant mass $s_0$. The contribution
of the $\Delta$-resonance is determined by the matrix element
\be
  <0|J_{\mu\,\alpha}^\Delta|\Delta(q,s)> &=& \lambda_\Delta
   \sqrt{\frac{2m_\Delta}{(2\pi)^3}}  u_{\mu\,\alpha}(q,s) e^{-iqx},
\ee
where $u_{\mu\,\alpha}(q,s)$ denotes a free Rarita-Schwinger
vector spinor normalized according to
\be
\sum_s u_{\mu\alpha}(q,s) \bar u_{\nu\beta}(q,s) = \left [
\frac {\hat q + m_\Delta}{2m_\Delta} \left(g_{\mu\nu} -
\frac 13 \gamma_\mu\gamma_\nu - \frac 23
\frac {q_\mu q_\nu}{m_\Delta^2} +
\frac 13 \frac{q_\mu\gamma_\nu-q_\nu\gamma_\mu}{m_\Delta}
\right)\right ]_{\alpha\beta}.
\ee
The short distance behavior of the coefficients of the spectral function
follows from matching the leading order singularities in the operator
product expansion of the correlator with the r.h.s. of eq. (5.5),
giving
\be
 a_1(q^2)=\frac{q^4}{160\pi^4},\hspace{1cm}
 a_2(q^2)=-\frac{5}{16} a_1(q^2).
\ee
Note that this result is indeed close to the pure spin 3/2 case
$a_2=-\frac 13 a_1$. Using (5.10) and (5.11) the invariant
correlation functions introduced above are given by
\setcounter{figure}{\value{equation}}
\renewcommand{\theequation}{5.12\alph{equation}}
\setcounter{equation}{0}
\be
  \Pi_1^\Delta(\tau) &=& \;\;\,2m_\Delta\lambda_\Delta^2 D(m_\Delta,\tau)
    +\frac{3|\langle\bar qq\rangle|}{4\pi^2}
    \int_{s_0}^\infty ds\, s D(\sqrt s,\tau), \\[0.1cm]
  \Pi_2^\Delta(\tau) &=& -2\lambda_\Delta^2 D^\prime (m_\Delta,\tau)
    -\frac{3}{256\pi^4}
    \int_{s_0}^\infty ds\, s^2 D^\prime (\sqrt s,\tau), \\[0.1cm]
  \Pi_3^\Delta(\tau) &=& -\frac{2\lambda_\Delta^2}{m_\Delta\tau}
     D^\prime (m_\Delta,\tau)
    -\frac{|\langle\bar qq\rangle|}{2\pi^2}
    \int_{s_0}^\infty ds\,\left(\frac1\tau D^\prime (\sqrt s,\tau)
     +\frac{s}{8} D(\sqrt s,\tau)\right)\! , \\[0.1cm]
  \Pi_4^\Delta(\tau) &=& \;\;\,2\lambda_\Delta^2 \left(
     -\frac{4}{m_\Delta^2\tau^2}D^\prime (m_\Delta,\tau)
      + \frac1\tau D(m_\Delta,\tau)\right)\nonumber \\[0.1cm]
      & & -\frac{3}{160\pi^4}
    \int_{s_0}^\infty ds\, \left( \frac{4s}{\tau^2}D^\prime (\sqrt s,\tau)
         - \frac{s^2}{\tau} D(\sqrt s,\tau)
         +\frac{s^2}{16}D^\prime (\sqrt s,\tau)
     \right) \! .
\ee
\renewcommand{\theequation}{5.\arabic{equation}}
\addtocounter{figure}{1}
\setcounter{equation}{\value{figure}}
\noindent
The results for the corelation functions are given in figure 4 and table 3.
The most important difference as compared to the nucleon
correlation function is the fact that the strong enhancement
over the free correlator which we observed in $\Pi_2^N$
and $\Pi_4^N$ is not present in $\Pi_2^\Delta$ and $\Pi_4^\Delta$.
This means that the $\Delta$ correlation function is qualitatively
different from the nucleon one. Note that this feature is not
reproduced by the vacuum dominance approximation. In particular,
vacuum dominance predicts a stronger enhancement in $\Pi_2^\Delta$
than in $\Pi_2^N$.

Fitting the parameters in (5.11) to the numerically
calculated correlators we find
\be
  m_\Delta = 1440\pm 70\,{\rm MeV},\hspace{1cm}
  \lambda_\Delta = 0.033\pm 0.005\,{\rm GeV}^3.
\ee
With these parameters, we obtain a very good description of {\it  all
four} correlation functions. The corresponding delta mass is
somewhat bigger than the experimental one. This situation  is similar
to the one encountered in the meson sector \cite{Shuryak_Ver_il2}:
the delta contains only vector
diquarks, and, like the  rho meson, it also
 turns out to be too heavy in the RILM.

As in the case of the nucleon one can speculate whether we really
observe a true bound state in the spin-isospin  3/2 channel or
whether the data can also be explained in terms of the propagation
of three essentially uncorrelated quarks. This question is actually
more serious in the case of the delta, since in this
correlator there are no strongly attractive diquark channels
present.  For that we reason we have compared the measured
delta correlation functions with the simple constituent model
introduced in the last section. The results are shown in figure 5.
Again, the contribution from three uncorrelated quarks clearly
underestimates the data. We conclude that although there is less
attraction in the delta channel as compared to the nucleon, the
RILM predicts a bound state in the spin-isospin 3/2 channel.

\vskip 1.5 cm
\centerline{\bf 6. Comparison with QCD sum rules and lattice
calculations}
\vskip 0.5 cm

   Predictions for baryonic correlators  based on QCD sum rules
were obtained by several groups. We show the latest results by
Belyaev and Ioffe \cite{Belyaev_Ioffe}, which are
compared to those from Chernyak et al.~(see \cite{FZOZ} and references
therein). The results of Chernyak et al.~are based on somewhat
different correlators and fitting procedures. In the nucleon case
the two predictions agree very well (and therefore we have plotted
only one of them), but they strongly disagree for the delta.

  Chernyak et al.~have also performed a more general calculation of 'baryonic
wave functions' and have concluded that  {\it there exists a qualitative
difference between the structure of octet and decuplet particles}.
Limited data on hard exclusive reactions involving  those particles seems
to confirm this important conclusion. Our results, based on the RILM, also
show that {\it the octet and the decuplet baryons have very different
correlation functions}. In particular the coupling constants (proportional
to the probability to find all three  quarks at the same point)
turn out to be  drastically different!

  Our results are compared to QCD sum rule predictions in Fig.6.
In the nucleon case the RILM results agree very well with all
sum rule predictions. In the delta channel this is not the
case: the predictions from  Chernyak et al.~are strongly favored. However,
we are unable to explain why we are in  agreement with certain sum
rule calculations but not with others. The whole topic is very confusing,
since (as it was demonstrated above in the simplest case of diquarks)
the agreement of OPE predictions with RILM results can
{\it depend on the  current used}, and for some quantities OPE results
based on different currents do not provide consistent answers.

  In general, it is hardly surprising that these two approaches do not
generally agree for many particular observables. In fact, the 'vacuum
dominance' hypothesis is based on the simplistic idea of a very smooth,
nearly homogeneous distribution of quark and gluon fields in
space time. On the contrary, the RILM is an example of a model with strongly
inhomogeneous fields, with gauge field concentrated in large fluctuations
of a particular structure, in this case small size instantons.

   Proceeding now to a comparison with lattice calculations, let us first
make some general comments. 'Traditional' lattice calculations use
plane-to-plane correlation functions\footnote{And sometimes even a specially
chosen 'smeared' sources in order to improve the  projection onto the
ground states.}, and those are known to have the following two problems:

\begin{itemize}
\item The measured ratio $m_N/m_\rho$ remains much closer to the 'naive quark
model' value 3/2 rather than the experimental one 1.2, even for the
lightest quarks studied (see e.g. \cite{Christ_etal}).

\item The lattice results for the nucleon-delta splitting appear
to be smaller than the experimental value  $m_\Delta-m_N=293$  MeV.
\end{itemize}

It was speculated  in \cite{Shuryak_cor}
that the nucleon, as measured on the lattice, seems to be
'too heavy', probably because {\it instanton-induced effects}
 were under-represented in those simulations, for whatever reason.

   In Fig.6 we compare our results with recent lattice data
\cite{Negele_etal}, which for the first time looked at point-to-point
correlators. Good agreement is observed, our RILM results literally
coincide with the lattice data  within  the error bars of the calculations!
This agreement is somewhat surprising, because
 our fit to  the RILM results gives $m_N/m_\rho = 1 \pm 0.1 $,
which is {\it smaller} than the experimental value whereas our prediction
for the splitting $m_\Delta-m_N=480$ MeV is {\it larger} than the
experimental result. Thus, deviations from experimental data in the RILM
results have {\it opposite} trends as compared to the deviations
observed in standard lattice gauge calculations.

Provided a fit to the Negele et al.~data gives resonance parameters similar
to the ones obtained by us, this particular set of lattice measurements seems
to show deviations from experimental data that are {\it opposite} as
compared to  other lattice works. One may speculate that the two sets of
lattice measurements emphasize the role of confinement and instanton-induced
forces differently, because plane-to-plane correlators deal with large
interquark distances, while point-to-point correlators are also sensitive
to smaller ones.

A controversial subject remains the question
whether octet and decuplet baryons have a qualitatively different
structure. Even a naive constituent quark model including a hyperfine
interaction leads to significantly different {\it mean square radii},
e.g. \cite{Isgur_Karl} $r_\Delta/r_N \approx 1.4$. If so, the probability
to find all three quarks at the same point should be about one order of
magnitude larger in the nucleon compared to the delta. However, lattice
studies \cite{Leinweber_etal} have been unable to find this effect: it was
concluded  that the nucleon and the delta have essentially {\it the same}
radius. This question certainly deserves further study.

  One more comment we would like to make concerns the large {\it finite size}
effects we have observed in our calculations. Going from 64 to 256 instantons,
(or from a box volume of $64\,{\rm fm}^4$ to $256\,{\rm fm}^4$) we have seen
significant changes in some correlation functions, the nucleon correlators
among them. Obviously, lattice measurements (which use significantly smaller
boxes) should be subject to similar modifications.

Clearly, point-to-point correlation functions are much more sensitive to the
dynamics of the interquark interactions at small distances. They are also more
stable with respect to finite size corrections. Further efforts in this
direction (using the TERAFLOP or similar next-generation projects)
should be encouraged.

\newpage
\vskip 1.5cm
\centerline{\bf 7. Discussion }
\vskip 0.5 cm

   Finally, having completed the series of three papers devoted to the RILM,
we would like to make  a few more general comments.

  Although the RILM is a model that was  originally designed only to describe
the most global properties of the QCD vacuum, the gluon and quark condensates,
it turns out to work very well. Not only does it reproduce small deviations
from asymptotic freedom at short distances, as it was originally anticipated,
but it correctly describes a large number of mesonic and baryonic correlators,
some of which vary over several  orders of magnitude. The results obtained
imply that even in the absence of confinement the lowest states in most
channels are in fact bound states of two or three quarks. If we translate
the measured correlators into resonance masses and coupling constants the
results agree with experimental data at the 10-15 percent level.

   Does this imply that by fixing the correct magnitude of two condensates,
one can reproduce all other vacuum parameters as a more or less direct
consequence? Absolutely not:
our experience with {\it interacting instanton ensembles} shows that even
inside the instanton framework it is very easy to get  drastically different
sets of correlators, which are completely incompatible with experiment, even
if the quark and gluon condensates are kept more or less fixed. Thus, the
excellent performance of the RILM is indeed a non-trivial
observation, which probably means it has a lot of truth in it.

  In any case, this model
definitely 'outperforms' such widely used theoretical approaches
as QCD sum rules or the Nambu-Jona-Lasinio model, and therefore applying
it to a broader set of questions appears
to be justified. Among the most fascinating topics to be studied
are nuclear forces at small distances (propagation of 6 quarks together),
hadronic structure and wave functions, multi-point correlators and many others.

  The central question posed at the beginning  of this paper has been
answered: the RILM describes the lightest baryons as some {\it bound} states.

 Moreover, their {\it spin splitting} comes out naturally,
without any need for additional assumptions or parameters.
  The splitting  between scalar and vector diquarks
is essentially due to the instanton-induced attraction present in
the scalar channel \cite{Betman_Laperashvili,Rosner_Shuryak}.
Although we have not considered hyperons in this work, the
correct dependence of the instanton-induced interaction
on the  strange quark mass \cite{Rosner_Shuryak} allows us to speculate
that these splittings will also come out reasonable.

  Furthermore,
 comparison between the calculated nucleon and delta correlators
provides another important lesson, namely: {\it the octet and the decuplet
baryons   have completely different wave functions}, because even the
coupling constants (proportional to the probability to find all three
quarks at the same point) are drastically different.
   Note that this picture of baryons is significantly different from
e.g. a 'skyrmion-based' one. According to the skyrmion  picture, a
nucleon and a delta are essentially the same object, just rotating with a
different angular (and isotopic) momentum.
According to our calculations, they are quite different already at
relatively small distances due to the strong instanton-induced attraction,
which is  present in octet baryons but absent in the decuplet.

\vskip 1.5 cm
{\bf \noindent 8. Conclusions \hfil}
\vglue 0.4cm

  We have performed a numerical analysis of diquark and baryonic
correlation functions in the framework of the random instanton
liquid model (RILM). Our main findings can be summarized as follows:

(i) The quark-quark interaction between light quarks is very much
channel-dependent, and it qualitatively resembles the one in the
quark-antiquark channels (with opposite parity). In particular,
similar to the situation in the pion and rho channels, the interaction
in the isospin I=0 (scalar diquarks) and I=1 (vector diquarks)
channels are completely different: the scalar channel has a much stronger
attractive interaction than the vector one, with the correlator reaching
a maximum value which is about an order of magnitude larger than
the free quark correlator.

(ii) If diquarks are treated as physical particles, the scalar and vector
diquarks are found to have masses of 420 and 940 MeV, respectively.
The non-relativistic description, related to  masses of heavy-light baryons
(minus the mass of the heavy quark), have  produced $\epsilon_{\Lambda_Q}=
760$ MeV and $\epsilon_{\Sigma_Q}=900-1000$ MeV, in reasonable agreement
with the limited experimental information and QCD sum rule calculations.

(iii) Calculation of all six nucleon and four delta  correlation functions
have been performed and analyzed. The octet (nucleon) and decuplet (delta)
correlators are found to be qualitatively different, with the main
difference being due to attractive interaction for scalar I=0 diquarks
mentioned above.

(iv) The nucleon parameters  agree well with the predictions from
different versions of the QCD sum rules, but in the case of the delta
the results from Chernyak et al.~(see \cite{FZOZ}) are strongly preferred.

(v) Our results were also found to be in agreement with the first lattice
data on point-to-point correlation functions \cite{Negele_etal}. Moreover, this
agreement is surprisingly good, the two calculations  practically agree inside
error bars.

\vskip 1.5 cm
{\bf \noindent 9. Acknowledgements \hfil}
\vglue 0.4cm

 The reported work was partially supported by the US DOE grant
DE-FG-88ER40388. We acknowledge the NERSC at Lawrence Livermore where
most of the computations presented in this paper were performed.
We are also indebted to J.~Negele, who has kindly supplied us with results
of his calculations prior to publication and has discussed them in detail.
\newpage
\centerline{\bf Figure Captions}
1. Diquark correlation functions in the RILM, measured as a function
of distance. All correlators are normalized to the perturbative results.
The diquark channels are labeled by the dirac matrix $\Gamma$
defining the current. The solid lines show a fit using the diquark model
discussed in section 2 and the dotted curves are the prediction from
the vacuum dominance model.

2. Correlation functions for $\Lambda$-type diquarks (a-b)
and $\Sigma$-type diquarks (c-d). The channels are labeled as in
figure 1. The solid lines now correspond to the heavy-light parametrization
discussed in section 3. The dotted and dashed lines in figure (a)
correspond to Grosin-Yakovlev prediction and their OPE expression,
respectively.

3. Nucleon correlation functions in the RILM. Channels are labeled as
in section 4 of the text. The solid line shows the 'nucleon pole plus
continuum' model, while the dotted curve is the vacuum dominance
approximation.

4. Delta Correlation functions in the RILM. Curves are as in figure 3.

5. Comparison of nucleon and delta correlation functions with simple
quark-diquark models. Figures (a) and (b) show the correlation functions
$\Pi_2^N$ (nucleon) and $\Pi_2^\Delta$ (delta), respectively.
Data and solid curves are as shown in figs.~3 and 4. The dotted
curves show the constituent model introduced in section 4 of
the text. In figure 5a, the quark-diquark model is given by the
dashed curve.

5. Comparison with QCD sum rules and lattice data for the two
correlation functions $\Pi_2^N$ (nucleon) and $\Pi_2^\Delta$
(delta). The RILM results are given by the solid triangles, the
lattice data from \cite{Negele_etal} by the open squares. QCD
sum rules predictions from Belyaev and Ioffe \cite{Belyaev_Ioffe}
are shown as long dashed lines, results from Chernyak et
al.~\cite{FZOZ} as short dashed lines.
\vfill
\newpage
\centerline{\bf Table Captions}
1. Numerical results from fitting the diquark correlation functions
in the RILM with a 'diquark resonance plus continuum' model. The
parameters are defined in section 2 of the text.

2. Results from fitting the diquark correlators using the non relativistic
parametrization related to the masses of heavy light baryons introduced
in section 3. The two different values for the $\Sigma_Q$ correspond
to the results obtained from the $\Gamma=\gamma_\mu$ and $\sigma_{\mu\nu}$
diquark channels. For comparison, we also quote the results from two
QCD sum rule calculations.

3. Nucleon and Delta parameters obtained from performing a global fit
to the six nucleon and four delta correlation functions defined in
section 4 and 5. Results are also compared to the experimental values
for the masses and QCD sum rule predictions for the coupling constants.
\vfill
\newpage

\vskip1cm
\begin{center}
\begin{tabular}{||l|c|c|c||}\hline\hline
     &   this work  &  other information  & comment  \\ \hline \hline
 $m_S$         &   $420\pm 30$ MeV    &   234 MeV    &
        NJL model \cite{Vogl} \\ \hline
 $m_{AV}$      &   $940\pm 20$ MeV    &   824 MeV    &
        NJL model \cite{Vogl}   \\ \hline
 $m_{T}$       &   $570\pm 20$ MeV    &              &            \\ \hline
 $g_S$         &   $0.225\pm 0.011\,{\rm GeV}^2$  &
                   $0.135\pm 0.025 \,{\rm GeV}^2$  &
        QCD sum rules \cite{Dosch_etal}     \\ \hline
 $g_{AV}$      &   $0.244\pm 0.010\,{\rm GeV}^2$     &      &       \\ \hline
 $g_T$         &   $0.134\pm 0.004\,{\rm GeV}^2$     &      &       \\
\hline\hline
\end{tabular}
\end{center}
\vskip1cm
\centerline{\bf\large table 1}
\vskip2cm
\begin{center}
\begin{tabular}{||l|c|c|c||}\hline\hline
     &   this work  &  other information  & comment  \\ \hline \hline
 $\epsilon_{\Lambda_Q}$  & $760\pm 30$ MeV &  $700\pm 150$ MeV    &
                  QCD sum rules  \cite{Shuryak_hl}        \\
                         &                 &  $780$ MeV &
                  QCD sum rules \cite{Grosin_etal}  \\ \hline
 $\epsilon_{\Sigma_Q}$   & $890\pm 30$ MeV & $1100\pm 200$ MeV    &
                  QCD sum rules \cite{Shuryak_hl}        \\
                         & $1005\pm 30$ MeV & $780$ MeV    &
                  QCD sum rules \cite{Grosin_etal}         \\ \hline
 $f_{\Lambda_Q}$         & $0.052\pm 0.005\,{\rm GeV}^3$  &
                     $0.02\,{\rm GeV}^3$   &
                  QCD sum rules  \cite{Shuryak_hl}    \\
                         &              & $0.0225\pm 0.0045\,{\rm GeV}^3$ &
                  QCD sum rules  \cite{Grosin_etal}  \\ \hline
 $f_{\Sigma_Q}$          & $0.011\pm 0.002\,{\rm GeV}^3$  &
                       $0.06\,{\rm GeV}^3$      &
                  QCD sum rules  \cite{Shuryak_hl}         \\
                         & $0.044\pm 0.008\,{\rm GeV}^3$ &
                           $0.0225\pm 0.0045\,{\rm GeV}^3$ &
                  QCD sum rules \cite{Grosin_etal} \\ \hline\hline
\end{tabular}
\end{center}
\vskip1cm
\centerline{\bf\large table 2}
\newpage
\vspace*{1cm}
\begin{center}
\begin{tabular}{||l|c|c|c||}\hline\hline
     &   this work  &  other information  & comment  \\ \hline \hline
 $m_N$         &  $960\pm30$ MeV      & 939 MeV     &    experiment  \\ \hline
 $\lambda_1$   &  $0.032\pm0.001\,{\rm GeV}^3$ &
                  $0.035\pm0.008\,{\rm GeV}^3$ &  QCD sum rules
                                \cite{Ioffe} \\ \hline
 $\lambda_2$   &  $0.080\pm0.004\,{\rm GeV}^3$ &
                                      &        \\ \hline
 $E_0$         &  $1920\pm 50$ MeV     &             &       \\ \hline\hline
 $m_\Delta$    &  $1440\pm 70$ MeV    & 1232 MeV     &  experiment  \\ \hline
 $\lambda_\Delta$ & $ 0.033\pm0.005\,{\rm GeV}^3$    &
                    $ 0.050\pm0.013\,{\rm GeV}^3$    &   QCD sum rules
                          \cite{Ioffe}    \\
               &                    &  $ 0.035 \,{\rm GeV}^3$ &
                 QCD sum rules  \cite{FZOZ} \\  \hline
 $E_0$         &  $1962\pm 101$ MeV &                &     \\ \hline\hline
\end{tabular}
\end{center}
\vspace*{1cm}
\centerline{\bf\large table 3}
\vfill
\eject

\end{document}